\documentstyle[sprocl]{article}

\font\eightrm=cmr8

\def\Journal#1#2#3#4{{#1} {\bf #2}, #3 (#4)}

\def\NCA{\em Nuovo Cimento}
\def\NIM{\em Nucl. Instrum. Methods}
\def\NIMA{{\em Nucl. Instrum. Methods} A}
\def\NPB{{\em Nucl. Phys.} B}
\def\PLB{{\em Phys. Lett.}  B}
\def\PRL{\em Phys. Rev. Lett.}
\def\PRD{{\em Phys. Rev.} D}
\def\ZPC{{\em Z. Phys.} C}

\def\st{\scriptstyle}
\def\sst{\scriptscriptstyle}
\def\mco{\multicolumn}
\def\epp{\epsilon^{\prime}}
\def\vep{\varepsilon}
\def\ra{\rightarrow}
\def\ppg{\pi^+\pi^-\gamma}
\def\vp{{\bf p}}
\def\ko{K^0}
\def\kb{\bar{K^0}}
\def\al{\alpha}
\def\ab{\bar{\alpha}}
\def\be{\begin{equation}}
\def\ee{\end{equation}}
\def\bea{\begin{eqnarray}}
\def\eea{\end{eqnarray}}
\def\CPbar{\hbox{{\rm CP}\hskip-1.80em{/}}}

\begin{document}

\title{MULTIPARTICLE PRODUCTION AT RHIC AND LHC: A CLASSICAL POINT OF VIEW}

\author{A. KRASNITZ}

\address{CENTRA and Faculdade de Ci\^encias e Tecnologia,\\
Universidade do Algarve, Campus de Gambelas,
P-8000 Faro, Portugal\\E-mail: krasnitz@ualg.pt}

\author{R. VENUGOPALAN}

\address{Department of Physics and RIKEN-BNL Research Center,\\
Brookhaven National Laboratory,
Upton, NY 11973, USA\\E-mail: raju@bnl.gov}

\maketitle\abstracts{ 
We report results of our ongoing 
nonperturbative numerical study of a classical effective theory 
describing low-$x$ partons in the central region of a heavy-ion
collision. In 
particular, we give estimates of the initial transverse energies and 
multiplicities for a wide range of collision regimes, including those at RHIC 
and at LHC.}

In the central-rapidity region of heavy-ion collisions
at RHIC and at LHC a
combination of very high center-of-mass energy with a very large
number of participating valence quarks is expected to lead to a
novel regime of QCD, one characterized by a very high parton
density. This regime may not be amenable to analysis by conventional methods,
such as multiple scattering or classical cascade descriptions, which ignore
the coherence of the gluon field~\cite{mscl}.

This coherence emerges naturally
from the classical effective field theory approach of McLerran and Venugopalan
(MV)~\cite{RajLar}.
If the parton density in the colliding nuclei is high at small $x$, classical
methods are valid. It has been shown recently that a RG-improved
generalization of this effective action reproduces several key results
in small-$x$ QCD~\cite{JKMW}.

The model can be summarized as follows. The high-$x$ and the low-$x$
partons are treated separately. The
former corresponds to valence quarks and hard sea partons. These
high-$x$ partons are considered recoilless sources of color charge.
For a large Lorentz-contracted nucleus, this results in a static
Gaussian distribution of their color charge density $\rho$ in the
transverse plane:
$$P\left([\rho]\right)\propto\, {\rm exp}\left[-{1\over{2\Lambda_s^2}}
\int{\rm d}^2r_t\rho^2(r_t)\right].$$ The variance $\Lambda_s$ of the
color charge distribution is the only dimensional parameter of the
model, apart from the linear size $L$ of the nucleus. For central
impact parameters, $\Lambda_s$ can be estimated in terms of single-nucleon 
structure
functions~\cite{GyulassyMclerran}.
It is assumed, in addition, that the
nucleus is infinitely thin in the longitudinal direction. Under this
simplifying assumption the resulting gauge fields are boost-invariant.

The small $x$ fields are then described by the classical Yang-Mills equations
\begin{equation}
D_\mu F_{\mu\nu}=J_\nu\label{eqmo}\end{equation}
with the random sources on the two light
cones:
$J_\nu=\sum_{1,2}\delta_{\nu,\pm}\delta(x_\mp)\rho_{1,2}(r_t).$
The two signs correspond to two possible directions of motion along the
beam axis $z$. As shown by Kovner, McLerran and Weigert (KMW)~\cite{KMW}, 
low-$x$ fields
in the central region of the collision obey sourceless Yang-Mills equations
(this region is in the forward light cone of both nuclei) with the initial
conditions in the $A_\tau=0$ gauge given by
\begin{equation}
A^i=A^i_1+A^i_2;\ \ \ \ A^\pm=\pm{{ig}\over 2}x^\pm[A^i_1,A^i_2].
\label{incond}\end{equation}
Here the pure gauge fields $A^i_{1,2}$ are solutions of (\ref{eqmo}) for each
of the two nuclei in the absence of the other nucleus.

\begin{figure}[hbt]
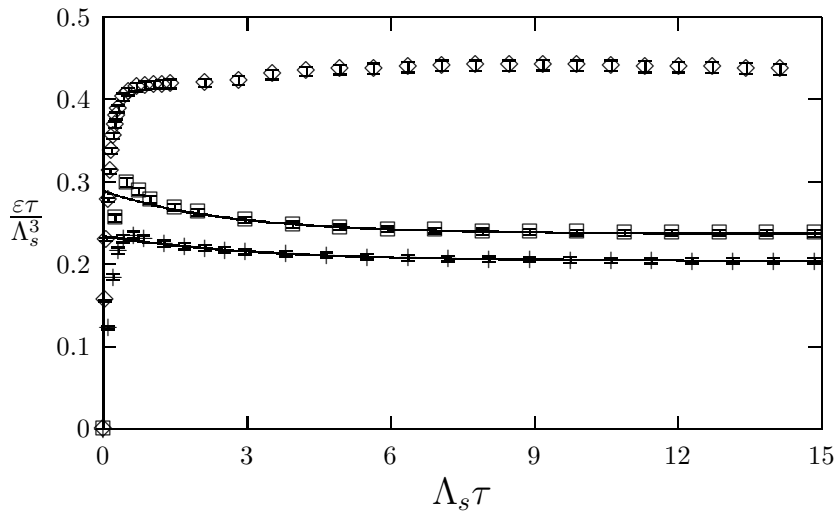

\setlength{\unitlength}{0.240900pt}
\ifx\plotpoint\undefined\newsavebox{\plotpoint}\fi
\sbox{\plotpoint}{\rule[-0.200pt]{0.400pt}{0.400pt}}%

\caption{Transverse-plane energy density per unit rapidity versus proper time
for the values 5.66 (diamonds), 35.36 (plusses), and 297 (squares) of 
$\Lambda_s L$.Both the energy density and the proper time are expressed in 
units of $\Lambda_s$.
The solid lines are fits of the data to the form
$\alpha+\beta\,{\rm exp}(-\gamma\tau)$.}
\label{ehist}\end{figure}

In order to obtain the resulting gluon field configuration at late
proper times, one needs to solve (\ref{eqmo}) with the initial condition 
(\ref{incond}).
Since the latter depends on the random color source,
averages over realizations of the source must be performed. 
KMW showed that in perturbation theory
the gluon number distribution by transverse momentum (per unit rapidity)
suffers from an	infrared divergence and argued that the distribution must
have the form
\begin{equation}
n_{k_\perp}\propto{1\over\alpha_s}\left({{\Lambda_s}\over k_\perp}\right)^4\ln\left({k_\perp\over{\Lambda_s}}\right) \label{dpt}\end{equation}
for $k_\perp\gg\Lambda_s$. 
The log term clearly indicates that the perturbative description breaks down
for $k_\perp\sim\Lambda_s$. 

A reliable way to go beyond perturbation theory is to re-formulate the EFT on
a lattice by discretizing the transverse plane. The resulting lattice theory
can then be solved numerically. We shall not dwell here on the details of the
lattice formulation, which is described in detail in Ref.~\cite{AlexRaj}.
Keeping in mind that
$\Lambda_s$ and the linear size $L$ of the nucleus
are the only physically interesting dimensional parameters of the
model~\cite{RajGavai}, we can write any dimensional quantity $q$ 
as $\Lambda_s^df_q(\Lambda_s L)$, where $d$ is the dimension of
$q$. All the non-trivial physical information is contained in the
dimensionless function $f_q(\Lambda_s L)$. 
We can estimate the values of the product
$\Lambda_s L$ which correspond to key collider experiments. Assuming
Au-Au collisions, we take $L=11.6$ fm (for a square nucleus!) and estimate
the standard
deviation $\Lambda_s$ to be 2 GeV for RHIC and 4 GeV for
LHC~\cite{GyulassyMclerran}.  Also, we
have approximately $g=2$ for energies of interest. The rough estimate is
then $\Lambda_s L\approx 120$ for RHIC and $\Lambda_s L\approx 240$ for LHC.

We now proceed to describe the results of our numerical study so far. Our 
simulations were performed for the SU(2) gauge group, in order to keep the
computational costs low. 

\begin{table}[h]
\begin{small}
\centerline{\begin{tabular}{|lrrrrrrr|} \hline
$\Lambda_s L$ & 17.68 & 35.36 & 70.7 & 106.06 & 148.49 & 212.13 & 296.98 \\
$f_E$ & $.323(4)$ & $.208(4)$
& $.200(5)$ & $.211(1)$ & $.232(1)$ & $.234(2)$ & $.257(5)$\\ \hline\end{tabular}}
\caption{The function $f_E$, {\it i.e.}, the energy per unit transverse area 
per unit rapidity, expressed in units of $\Lambda_s$, tabulated 
vs $\Lambda_s L$.} 
\end{small}
\end{table}

We first compute the energy per unit transverse area  per unit rapidity,
deposited in the central region by the colliding nuclei. As Figure \ref{ehist}
illustrates, this quantity tends to a constant at late proper times. 
We find that at $\tau\longrightarrow\infty$ the energy density in units of 
$\Lambda_s$
depends on the dimensionless parameter $\Lambda_s L$ as described in 
Table 1.
Note the very slow variation of this dimensionless function in
the entire range of $\Lambda_s L$ values, which includes both our RHIC and LHC
estimates. Using this plot, and assuming, in accordance with
Ref.~\cite{Muell2},
the $(N_c^2-1)/N_c$ dependence of the energy on the number of colors $N_c$,
we arrive at the values of 2700 GeV and of 25000 GeV for the transverse energy
per unit rapidity at RHIC and at LHC, respectively~\cite{AlexRaj1}.

\begin{figure}[hbt]
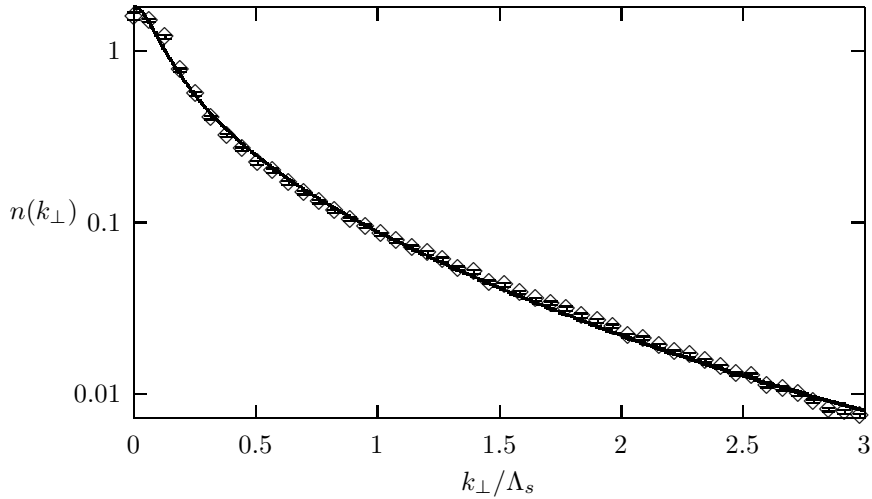

\setlength{\unitlength}{0.240900pt}
\ifx\plotpoint\undefined\newsavebox{\plotpoint}\fi
\sbox{\plotpoint}{\rule[-0.200pt]{0.400pt}{0.400pt}}%

\caption{$n(k_\perp)\equiv dN/L^2/d^2 k_\perp$ as a function of the
gluon momentum $k_\perp$ for $\Lambda_s L=297$ (diamonds).
The solid line is a fit to Bose-Einstein distribution, corresponding to a gas
of free particles at the inverse temperature $\beta=(1.28\pm 0.01)\Lambda_s$
and with the mass $m=(0.092\pm 0.003)\Lambda_s$.}

\label{nvsk}\end{figure}

The number and distribution of produced gluons
are of considerable interest as initial data for possible
evolution of the gluon gas towards thermal equilibrium~\cite{Muell2,evol,BMSS}.
Strictly speaking, a particle number is
only well-defined in a free-field theory, and 
there is no unique extension of this notion to a general interacting case.
For this reason we use two different
generalizations of the particle number to an interacting theory, each having
the correct free-field limit. We verify that the two definitions agree in
the weak-coupling regime corresponding to late proper times in the central
region~\cite{RVAK}. 

Our first definition is straightforward. We impose the Coulomb gauge condition
in the transverse plane: ${\vec\nabla}_\perp\cdot{\vec A}_\perp=0$ and
and determine the momentum components of the resulting field configuration.
Our second definition is based on the behavior of a free-field theory under
relaxation. Consider a simple relaxation equation for a field in real space,
\begin{equation}\partial_t\phi(x)=-\partial V/\partial\phi(x),\label{relax}
\end{equation}
where $t$ is the relaxation
time (not to be confused with real or proper time) and $V$ is the
potential. 
It is then easy to derive the following integral expression
for the total particle number of a free-field system:
\begin{equation}
N=\sqrt{8\over\pi}\int_0^\infty{{{\rm d}t}\over\sqrt{t}}\,V(t),\label{ncool}
\end{equation}
where $V(t)$ is the potential for the relaxed field.
Now (\ref{relax}) can be solved numerically for interacting fields.
Subsequently, $V(t)$ can be determined, and $N$ can be computed by numerical
integration. 
This
technique presently only permits determination of the total particle number.

We now present our results using both the techniques discussed. 
Our results for 
the number distribution, computed in
Coulomb gauge, are as follows.
We have verified that for large $k_\perp$ our numerically obtained multiplicity
agrees with the lattice analogue of the perturbative expression (\ref{dpt}).
At smaller $k_\perp$, the distribution softens
and converges to a constant value, unlike its perturbative counterpart.  
Notably, this qualitative change of the distribution occurs at 
$k_\perp\sim\Lambda_s$. We tried to quantify this non-perturbative
behavior by fitting the distribution to a variety of physically motivated
functional forms.
Surprisingly, we find that the shape of the distribution 
is closely reproduced by the Bose-Einstein form 
$n(k)=A/(\exp(\beta\omega_k)-1)$,
with the inverse temperature $\beta$ of the order of 1 in units of $\Lambda_s$,
and with $\omega_k$ corresponding to a free massive dispersion relation, with
the mass of the order of $0.1\Lambda_s$.
This is an unexpected result for a purely classical theory, whose 
meaning, beyond providing us with a convenient parametrization, 
is not yet clear.

Table 2 summarizes our findings for the total gluon number, written as 
$N=(\Lambda_s L/g)^2 f_N(\Lambda_s L)$. Note the good agreement between the two
methods used to determine $N$ (a more detailed analysis shows that the small
difference between the two is likely an artifact of the Coulomb gauge). 
Note that, similarly to $f_E(\Lambda_s L)$, $f_N(\Lambda_s L)$ is a 
slowly-varying function. Based on Table 2, and making a naive extrapolation
from the SU(2) to the SU(3) gauge group, we estimate the initial gluon
number per unit rapidity as $\sim 950$ for central Au-Au collisions at 
RHIC energies, and as $\sim 4300$ for central Au-Au collisions at LHC.
Let us again emphasize that, even though our figure for RHIC is not far from the
recent experimental result, both the distribution and the total multiplicity
are likely to be changed by further evolution of the field configuration.

\begin{table}[hbt]
\begin{small}
\centerline{\begin{tabular}{|lrrrrrr|} \hline
$\Lambda_s L$ & 35.36 & 70.71 & 106.1 & 148.5 & 212.1 & 297.0 \\
$f_N$ (cooling)	 & .116(1) & .119(1) & .127(1) & .138(1) &
.146(1) & .151(1)  \\
$f_N$ (Coulomb) & .127(2)  & .125(2) & .135(1) & .142(1) & .145(1)  & .153(1) \\
\hline
\end{tabular}}
\caption{Values of $f_N$ vs $\Lambda_s L$ for the two definitions of
the total particle number.}
\label{nvsmultab}
\end{small}
\end{table}

At the current stage of our project we are only able to make qualitative 
predictions. One obvious way to improve the accuracy is by 
replacing the SU(2) color group by the physical SU(3) one. Another is by 
allowing deviations from the strict boost invariance. These issues will be 
addressed in the future.

\section*{Acknowledgments}
We would like to thank Larry McLerran and Al Mueller for very useful
discussions.  R. V.'s research was supported by DOE Contract
No. DE-AC02-98CH10886.	The authors acknowledge support from the
Portuguese FCT, under grants CERN/P/FIS/1203/98 and
CERN/P/FIS/15196/1999.


\section*{References}
\begin{thebibliography}{99}
\bibitem{mscl}
X.-N. Wang, {\em Phys. Rep.} {\bf 280} 287 (1997);
K. Geiger, {\em Phys.Rep.} {\bf 258} 237 (1995)
B. Zhang, {\em Comput. Phys.Commun.} {\bf 104} (1997) 70.
\bibitem{RajLar}
L. McLerran and R. Venugopalan, {\em Phys. Rev.} {\bf D49} 2233 (1994);
{\bf D49} 3352 (1994); {\bf D50} 2225 (1994).
\bibitem{JKMW}
J. Jalilian--Marian, A. Kovner, L. McLerran, and H. Weigert, {\em Phys. Rev.}
{\bf D55} (1997) 5414;
J. Jalilian-Marian, A. Kovner, A. Leonidov, and H. Weigert,
{\em Nucl. Phys.} {\bf B504} 415 (1997); {\em Phys. Rev.} {\bf D59}
034007 (1999); Erratum-{\it ibid.} {\bf D59} 099903 (1999);
J. Jalilian-Marian, A. Kovner, and H. Weigert, {\em
Phys. Rev.} {\bf D59} 014015 (1999); L. McLerran and R. Venugopalan, {\em
Phys. Rev.} {\bf D59} 094002 (1999); Yu.V. Kovchegov, Phys.\ Rev.\ {\bf D 54} 
(1996) 5463, hep-ph/9605446.
\bibitem{GyulassyMclerran} M. Gyulassy and L. McLerran, {\em
Phys. Rev.}  {\bf C56} (1997) 2219.
\bibitem{KMW}
A. Kovner, L. McLerran and H. Weigert, {\em Phys. Rev} {\bf D52} 3809
(1995); {\bf D52} 6231 (1995).
\bibitem{AlexRaj}
A. Krasnitz and R. Venugopalan, hep-ph/9706329, hep-ph/9808332;
{\em Nucl. Phys.} {\bf B557} 237 (1999).
\bibitem{RajGavai}R. V. Gavai and R. Venugopalan, {\em Phys. Rev.} {\bf D54}
5795 (1996).
\bibitem{Muell2}
A. H. Mueller, Nucl. Phys. {\bf B572} (2000) 227, hep-ph/9906322; 
Phys. Lett. {\bf B475} (2000) 220, hep-ph/9909388.
\bibitem{AlexRaj1}
A. Krasnitz and R. Venugopalan, Phys. Rev. Lett. {\bf 84} (2000) 4309, 
hep-ph/9909203; hep-ph/9910391.
\bibitem{evol}J. Bjoraker and R. Venugopalan, Phys. Rev. {\bf C63} (2001) 
024609, hep-ph/0008294; A. Dumitru and M. Gyulassy, hep-ph/0006257.
\bibitem{BMSS} R. Baier, A.H. Mueller, D. Schiff, D.T. Son, hep-ph/0009237.
\bibitem{RVAK} A. Krasnitz and R. Venugopalan, hep-ph/0007108.
\end{thebibliography}
\end{document}